\documentclass[fleqn,twoside]{article}
\usepackage{espcrc2}

\usepackage{graphicx}
\usepackage[figuresright]{rotating}

\newcommand{\AmS}{{\protect\the\textfont2
  A\kern-.1667em\lower.5ex\hbox{M}\kern-.125emS}}

\title{Full QCD Algorithms towards the Chiral Limit}

\author{M. Hasenbusch \address[MCSD]{NIC/DESY Zeuthen,  
          Platanenallee 6, D-15738 Zeuthen, Germany} }
       
\begin{document}

\begin{abstract}
I discuss the behaviour of algorithms for dynamical fermions as 
the sea-quark mass decreases. I focus on the Hybrid-Monte-Carlo (HMC)
algorithm
applied to two degenerate flavours of Wilson fermions. First, I briefly
review the performance obtained in large scale HMC simulations.
Then I discuss a modified pseudo-fermion action for the HMC simulation
that has been introduced three years ago.
I summarize 
recent results obtained with this pseudo-fermion action  by the QCDSF and the 
ALPHA collaborations.
I  comment on alternatives to the HMC, like 
the Multiboson algorithm and variants
of it.
\vspace{1pc}
\end{abstract}

\maketitle

\section{INTRODUCTION}
Today it is understood that further progress in the simulation of lattice 
QCD requires dynamical fermions. First large scale simulations with 
two flavours of degenerate
Wilson fermions are done at a rather coarse lattice spacing $a \approx 0.1$
and are restricted to $m_{PS}/m_{V} > 0.5$. 
It is doubtful whether $\chi$-perturbation theory  can close the 
gap to the physical value $m_{\pi}/m_{\rho} \approx  0.18$ \cite{panel2002}.
It seems that, for a given numerical effort, Kogut-Susskind (KS) fermions 
allow to reach smaller sea-quark masses than Wilson fermions \cite{Gottlieb}.
For a critical review see \cite{Karl}. Yet,
biased by my personal experience, I shall restrict the following 
discussion to Wilson fermions. I expect that some statements
also apply to KS fermions.
To assess the progress that can be made with the machines of the near future,
like QCDOC or apeNEXT, 
we have to understand the scaling of the algorithms 
at hand. Also it is clear that we should not only rely on the 
speed-up of computers but also should work on the simulation algorithms. 
Recent progress \cite{MHschwinger,Peardon} with the Hybrid-Monte-Carlo (HMC) 
algorithm \cite{HMC} and new ideas \cite{Luescher} presented earlier this year
show that it is worth while to pursue this direction.
For simplicity, we shall focus in the following on two
degenerate flavours of sea-quarks. The corresponding 
partition function is given by 
$
Z =\int \mbox{D}[U] \;\; \exp(-S_G[U]) \; \mbox{det} M[U]^2 \;,
$
where $S_G[U]$ is the gauge action and $M[U]$ the fermion matrix. For
the definition 
see textbooks. 
Since the explicit calculation of the determinant of the fermion matrix
is not feasible for reasonably large lattices, so called pseudo-fermion 
fields are introduced \cite{WePe}:
\begin{eqnarray}
\label{pseudofermion}
&&\mbox{det} M[U]^2 \propto \;
\nonumber \\
&& \int \mbox{D}[\phi] \int \mbox{D}[\phi^{\dag}]
    \exp(-S_{PF}[U,\phi,\phi^{\dag}]) \;\;,
\end{eqnarray}
where $S_{PF}=|M[U]^{-1} \phi|^2$ is the pseudo-fermion action.
Since the inverse of the fermion matrix appears, the pseudo-fermion  action
$S_{PF}$
is non-local and hence a local up-date of the gauge-field 
requires $O(Volume)$ operations,
where $Volume$ is the number of sites of the lattice. It follows that
a full sweep over the lattice costs $O(Volume^2)$ operations.
To avoid this unfavourable scaling of the costs with the $Volume$, 
molecular dynamics evolution of the gauge field and eventually the
HMC algorithm were introduced. 

Later the multi-boson algorithm \cite{MultiB} was introduced. The 
inverse of the fermion matrix is approximated by some polynomial. 
This polynomial is factorized into its roots and for each factor, 
a boson-field is introduced. This way, the effective action becomes 
local. It can be simulated with a standard local  
Monte Carlo algorithm. However, it turned out that autocorrelation times 
increase considerably with increasing order of the polynomial.


\section{THE HMC ALGORITHM}
To allow the 
collective molecular dynamics evolution of the gauge-fields, new 
auxiliary variables, 
traceless Hermitian momenta $P_{x,\mu}$ conjugate to the gauge-field 
are introduced.
The resulting effective Hamiltonian consists of the gauge action, 
the pseudo-fermion action and a new part for $P$: $H[U,\phi,\phi^{\dag},P] = $
$S_G(U) + S_{PF}[U,\phi,\phi^{\dag}] +
\frac{1}{2} \sum_{x,\mu} \mbox{Tr} P_{x,\mu}^2$ . 
One elementary update (``trajectory'') of the HMC algorithm is composed
of the following steps:

a) Global heat-bath of the pseudo-fermions and the conjugate momenta. 

b) Molecular dynamics evolution of the gauge-field $U$ and the conjugate
      momenta $P$ with fixed pseudo-fermions $\phi$. 

c) Accept/Reject step: the gauge-field $U'$ that is generated by the
      molecular dynamics evolution is accepted with the probability 
$P_{acc}= $
$\mbox{min} [1, \exp(-H[U',\phi,\phi^{\dag},P']+H(U,\phi,\phi^{\dag},P])]$,
where $P'$ represents the conjugate momenta generated in the
molecular dynamics evolution. \\
The last step is needed, since
the molecular dynamics evolution can not be done exactly and hence
requires a numerical integration scheme,
resulting in 
$\Delta H=H[U',\phi,\phi^{\dag},P']-H(U,\phi,\phi^{\dag},P])] \ne 0$.

\subsection{Integration Schemes}
\label{schemes}
The HMC algorithm requires that the integration scheme used is 
area preserving and reversible. The simplest scheme that fulfills
these requirements is the so called ``leap-frog'' algorithm. 
Let us define the update of the gauge-field and the momenta as
\begin{eqnarray}
 T_U (\delta \tau)  &:& \; U \rightarrow e^{i \delta \tau \; P }
                         \; U \nonumber \\
 T_P (\delta \tau)  &:& \;  P \rightarrow P
  - i \delta \tau \; 
                      \delta (S_g[U] + S_{PF}[U])  \;. \nonumber
\end{eqnarray}
Then, the elementary step of the leap-frog algorithm is given by
\begin{equation}
\label{ourleap}
 T_2(\delta \tau) = T_P \left(\frac{\delta \tau}{2} \right) \; T_U(\delta \tau) \;
                    T_P \left(\frac{\delta \tau}{2} \right) \;\;\;.
\end{equation}
A trajectory is composed of $N_{md}$
consecutive  elementary steps.
Its integration error is $O(\delta \tau^2)$ for a trajectory of a given
length, where $\delta \tau$ is the step-size.  
Note that the order of the updates of momenta and gauge-fields is not unique.
In fact, in ref. \cite{JLQCD} it was demonstrated that the alternative order
$ T_2'(\delta \tau) = T_U \left(\frac{\delta \tau}{2} \right) \; T_P(\delta \tau) \;
                    T_U \left(\frac{\delta \tau}{2} \right) \;$
achieves the same acceptance rate as eq.~(\ref{ourleap})
(see fig. 2 of ref. \cite{JLQCD}) with  a roughly $15 \%$ larger step size
$\delta \tau$.
In the literature \cite{SeWe,CrGo,CaRo}
also higher-order schemes are discussed. These schemes
become increasingly complicated as the order increases. Recent studies,
e.g. \cite{Takaishi,CPPACS},
with the standard pseudo-fermion action~(\ref{pseudofermion}) 
show that higher-order schemes become less efficient than the
simple leap-frog integration scheme
as the sea-quark mass decreases.

Sexton and Weingarten \cite{SeWe} also suggested 
to split the update of the momenta into two
parts: 
\begin{eqnarray}
T_{PG} (\delta \tau) \; &:&  P \rightarrow P
- i \delta \tau \; 
\delta_U S_g[U] \nonumber  \\
T_{PF} (\delta \tau) \; &:&  P \rightarrow P
 - i \delta \tau \; 
\delta_U  S_{PF}[U] \;\;. \nonumber  
\end{eqnarray}
The leap-frog scheme is now generalized to
\begin{eqnarray}
\label{ts}
T_2(n,\delta \tau) \;=\;  T_{PF} \left( \frac{\delta \tau}{2} \right)
\;\;\;\;\;\;\;\;\;\;
\;\;\;\;\;\;\;\;\;\;\;\;\;\;\;\;\;\;\;\;\;\;\;\; &&  \\
\left[
T_{PG} \left(\frac{ \delta \tau}{2 n} \right)
T_U   \left(\frac{ \delta \tau}{n} \right)
T_{PG} \left(\frac{ \delta \tau}{2 n} \right)
\right]^n
T_{PF} \left( \frac{\delta \tau}{2} \right) && \nonumber
\end{eqnarray}
This allows to put a part of the action that is simple to compute (e.g. $S_G$)
on a smaller time-scale than the rest. 


\subsection{Preconditioning}
It has been noticed for a while that preconditioning of the fermion
matrix is indispensable in lattice QCD simulations. First, the number 
of iterations needed to solve $M^{-1} \phi$ is reduced, but also 
the step-size of the HMC algorithm can be increased (See e.g. \cite{ourbench}). 
In addition to the even-odd preconditioning, SSOR preconditioning and
variants of it \cite{Wupp} have been used in practice. 
The authors of ref. \cite{JaLi} have detailed how even-odd preconditioning
can be applied to clover-improved Wilson fermions. They propose two possible
variants.
The asymmetric version:
\begin{equation}
\label{asy}
\mbox{det} M \propto
 \mbox{det} (1_{ee} + T_{ee}) \;
                     \mbox{det} \hat M \;,
\end{equation}
where
$\hat M = 1_{oo} + T_{oo}  -
         H_{oe} (1_{ee} + T_{ee})^{-1} H_{eo}$.
$H_{oe}$ and $H_{eo}$ is the hopping-matrix, connecting odd with even
sites and vice versa. $T_{oo}$ and $T_{ee}$ encode the clover term.
The symmetric version is
\begin{equation}
\label{symm}
\mbox{det} M \propto \mbox{det} (1_{ee} + T_{ee})
                     \mbox{det} (1_{oo} + T_{oo})
                     \mbox{det} \hat M_{sym} \;,
\end{equation}
where now $\hat M_{sym} = 1_{oo}  -
(1_{oo} + T_{oo})^{-1}
M_{oe} (1_{ee} + T_{ee})^{-1} M_{eo}$. In most simulations, 
the asymmetric version~(\ref{asy}) has been used. However, recently
ref. \cite{JLQCD} reported that the symmetric version~(\ref{symm}) allows for 
a considerably larger step-size of the leap-frog than (\ref{asy}).

\section{SCALING OF THE HMC}
At the Lattice 2001 T. Lippert \cite{Lippert} reported about the scaling 
of the HMC algorithm based on the simulations of the SESAM collaboration.
The result can be summarized as  
\begin{eqnarray}
N_{flops} &=& 2.3(7) \cdot 10^7 \cdot \left(\frac{L}{a} \right)^5
             \cdot \left(\frac{1}{a m_{PS}} \right)^{2.8(2)} \nonumber \\
N_{flops} &=& 1.6(4) \cdot 10^7 \cdot \left(\frac{L}{a} \right)^5
             \cdot \left(\frac{1}{a m_{PS}} \right)^{4.3(2)} \nonumber 
\end{eqnarray}
for $\beta=5.6$ and $5.5$, respectively.
The increase of the costs with decreasing mass $m_{PS}$ of the 
pseudo-scalar meson has three sources: \\

a) Since the condition number of the fermion matrix increases as $m_{PS}$
decreases, the solver needs more iterations to compute $M[U]^{-1} \phi$.

b) Decrease of the step-size  $\delta \tau$ of the leap-frog integration scheme.

c)
Increase of the autocorrelation time. \\

Here I like to briefly discuss the  behaviour in two recent large scale 
simulations using clover-improved Wilson fermions.
The CP-PACS collaboration \cite{CPPACS} has simulated the Iwasaki gauge action
at three $\beta$-values corresponding to the lattice spacings  
$a\approx 0.22, 0.16$ and $0.11$ fm.
The JLQCD collaboration \cite{JLQCD} collaboration simulated 
the Wilson gauge action with $\beta=5.2$ corresponding to $a \approx 0.09$ fm. 

In table 
VI of ref. \cite{CPPACS} integrated auto-correlation times for the 
plaquette, the number of iterations $N_{inv}$ of the BiCGStab solver
and the effective pion mass are given. Among these, $\tau_{Ninv}$ is the 
largest. For $\beta=2.1$ ($a=0.11$ fm) it behaves as 
$\tau_{Ninv} \propto (\kappa_c - \kappa)^{-0.3(2)}$. At the two smaller
values of $\beta$, we see a slightly larger increase of the auto-correlation 
times, corresponding to the exponents $-0.5(1)$ and $-0.4(1)$. JLQCD 
\cite{JLQCD} report autocorrelation times in table IV. In their case, 
the autocorrelation time of the plaquette even decreases as $\kappa$
increases. 
Next, let us discuss the number of iterations required by the BiCGStab solver.
The numbers for the $20^3 \times 48$ lattice given in table I of ref. 
\cite{JLQCD} can be fitted as
\begin{equation}
N_{inv} = 0.253(4) \;\;  (\kappa_c-\kappa)^{-0.893(2)} \;\;.
\end{equation}
The numbers for $N_{inv}$ given in table II of ref. \cite{CPPACS}  lead to 
virtually the same exponent. Fitting these numbers as a function function
of $m_{PS}$, taken from table XXIII of ref. \cite{CPPACS} gives
$N_{inv} \propto  m_{PS}^{-1.69}$.
Finally, let us consider the step-size of the integration scheme.
Taking the numbers for $\beta=2.1$ from table II of ref. \cite{CPPACS}
for the step-size, 
we find $\delta \tau \propto (\kappa_c-\kappa)^{0.6}$. It is more difficult
to extract such a result from the numbers of ref. \cite{JLQCD}, since the 
acceptance rate for the different values of $\kappa$ vary quite a bit.
Nevertheless it is clear from the numbers that the step-size has to be 
reduced considerably as $\kappa$ increases. 
Taking into account the increase of $N_{inv}$ and the decrease of 
$\delta \tau$ we get $cost \propto m_{PS}^{-z}$ with $z\approx 3$, 
which is similar to 
the results of ref. \cite{Lippert}.

%

At the Lattice 2002, A. Irving \cite{Irving} and 
Y. Namekawa \cite{Namekawa} reported about explorative studies 
at light quark masses with $\frac{m_{PS}}{m_V} \approx 0.4$. Details of 
ref. \cite{Irving} will be given below. In ref. \cite{Namekawa} the Iwasaki
gauge action at $\beta=1.8$ ($a\approx 0.22$ fm) was studied 
at $\kappa$-values corresponding to $m_{PS}/m_V=0.6$, $0.5$ and $0.4$. 
They find that the BiCGStab algorithm frequently fails to converge at the 
smaller quark masses. This problem was overcome by replacing the 
BiCGStab with the BiCGStab(DS-L) algorithm \cite{DSL},
which is a generalization of the BiCGStab. 
From fig. 2 of ref. \cite{Namekawa}
we see that a step size of $\delta \tau=0.003$ had to be
used at $m_{PS}/m_V \approx 0.5$ for a $12^3 \times 24$ lattice.
Despite of this small step-size, spikes (i.e. extremely large $\Delta H$)
frequently occur (See fig. 2
of ref. \cite{Namekawa}). 
They find that a reduction of the step-size $\delta \tau$ rapidly reduces the 
frequency of these spikes.

\section{MODIFIED PSEUDO-FERMIONS}
It has been observed by various authors that replacing the original fermion 
matrix  by the even-odd preconditioned one
allows for a larger step-size $\delta \tau$ in the HMC at
constant acceptance rate. This lead to the idea \cite{MHschwinger} to 
factorized the fermion matrix such that
each part has a smaller condition number than the original fermion matrix. 
In the modified pseudo-fermion action,
for each of the factors a pseudo-fermion field is introduced.
The splitting of the fermion matrix $M$ can be written as
$\bar M  \;\; := \;\;  W^{-1} \; M \;$.
In ref. \cite{MHschwinger} $W$ has been chosen as fermion matrix with 
a smaller hopping-parameter than $M$ itself. Equivalently we get
\begin{equation}
\label{original}
 W = M + \rho \;\;.
\end{equation}
A second choice that we studied \cite{HaJa} is inspired by twisted mass QCD
\begin{equation}
\label{rainer}
 W = M + i \rho  \gamma_5 \;\;.
\end{equation}
We introduce pseudo-fermions for both $W$
and $\bar M$:
\begin{eqnarray}
&&\mbox{det} M M^{\dag} \propto
\int {\cal D} \phi_1^{\dag} \int {\cal D} \phi_1 \int {\cal D}\phi_2^{\dag}
\int {\cal D} \phi_2 \nonumber \\
&& \;\;\;\;\; \;\;\; \exp(-|W^{-1} \phi_1|^2\;-|\bar M^{-1} \phi_2|^2)
\end{eqnarray}
Hence, the modified pseudo-fermion action is then given by
\begin{equation}
\label{modified}
 S_F = S_{F1} + S_{F2} = |W^{-1} \phi_1|^2\;+|\bar M^{-1} \phi_2|^2 \;\;.
\end{equation}
For the practicability of the HMC algorithm it is important that the 
variation of $S_F$ can be easily computed. For our choices of $W$,
the variation of $S_{F1}$ can
be computed in the same way as the variation of $S_F$ of 
eq.~(\ref{pseudofermion}).
Also the 
variation of $S_{F2}$ can be explicitely computed: $\;\;\;\;\delta S_{F2} = $
\begin{equation}
 X^{\dag} \delta M Y - Y^{\dag} \delta M^{\dag} X
                + X^{\dag} \delta W \phi_2  +\phi_2^{\dag} \delta W^{\dag} X
\end{equation}
where
$ X = M^{\dag -1} M^{-1} W \phi_2 ,\; Y = M^{-1} W \phi_2$.
The experience with the 2D Schwinger model \cite{MHschwinger} has shown
that the step-size of the integration scheme can indeed be increased
by replacing the standard by the modified pseudo-fermion action.
For the largest value of $\kappa$ that we have studied, the step-size could
be increased by a factor of two, while keeping the acceptance rate fixed. 
For a related approach see ref. \cite{Peardon}.

\subsection{Numerical results for Lattice QCD}
Recently, detailed studies \cite{HaJa,QCDSF,ALPHA} 
of the modified pseudo-fermion action applied
to the HMC simulation of Lattice QCD with two flavours of dynamical Wilson
fermions have been carried out.  
In these studies, the method has been
applied on top of
even-odd preconditioning. I.e. the splitting is applied to $\hat M$.
All three studies have used the asymmetric version~(\ref{asy}) and
(unfortunately) not eq.~(\ref{symm}) as recommended by ref. \cite{JLQCD}.
The first question is, whether also here the step-size can 
be increased and how this scales with the sea-quark mass and the lattice size.
In addition, it is important to check, whether there are
effects of the modified pseudo-fermion action on the number of iterations
needed by the solver, the autocorrelation times
and the reversibility of the integration scheme.

In ref. \cite{HaJa} we simulated at $\beta=5.2$ with
$c_{sw}= 1.76$. Note that the final result of 
ref. \cite{alphacsw} is $c_{sw} = 2.0171$. 
First we studied a
$8^3 \times 24$ lattice at $\kappa=0.137$.
Later we simulated a $16^3 \times 24$ lattice at
$\kappa=0.139$, $0.1395$ and $0.1398$. Following 
ref. \cite{thesis}, these values of $\kappa$
correspond to $m_{PS}/m_{V} \approx 0.856, 0.792, 0.715$ and $0.686$,
respectively.

We have tested two different integration schemes: The standard leap-frog
and a partially improved one suggested by Sexton and Weingarten
(see eq.~(6.4) of ref. \cite{SeWe})
that has a reduced amplitude of the  $O(\delta \tau^2)$ error
compared with the leap-frog scheme.
We used a trajectory length 1 throughout. To find the 
optimal value of the parameter $\rho$ of the modified 
pseudo-fermion action, we have performed 
runs for different values of $\rho$ with $100$ trajectories each.  
For all these runs we used the same step-size $\delta \tau$. We looked
for the value of $\rho$ that gives the maximal acceptance rate. This 
search is not too difficult, since there seems to be only one local 
maximum, which is rather broad.

We have tested both splittings eq.~(\ref{original}) and  eq.~(\ref{rainer}).
We found a small advantage for eq.~(\ref{original}). 
A very important result is that the partially improved scheme gains more 
from the modified pseudo-fermion action than the leap-frog scheme. 
In particular for our smallest quark mass, given by $m_{PS}/m_{V} \approx
0.686$, the leap-frog performs better than the partially improved scheme 
in the case of the standard pseudo-fermion action, while this order is just 
reversed when the modified pseudo-fermion action is used. In particular, 
we find for the $16^3 \times 24$ lattice with the 
standard pseudo-fermion action and the leap-frog scheme that
$\delta \tau=0.01$ is needed to get $P_{acc}=0.77(3)$.  With the 
modified  pseudo-fermion action at the optimal $\rho$ 
already $\delta \tau=0.02$ gives $P_{acc}=0.76(2)$.
With the partially improved integration scheme and the standard 
pseudo-fermion action we get with 
$\delta \tau=0.0166...$ an acceptance rate $P_{acc}=0.82(2)$.
The modified pseudo-fermion action with $\delta \tau=0.066...$ gives
$P_{acc}=0.74(3)$. Comparing the two integration schemes, we have to note 
that for one elementary step, for the partially improved scheme, 
the variation of the force has to be computed 
twice as often as for the leap-frog. It remains an advantage of
about $(0.066...)/(2 * 0.01)=3.33 ...$ for the partially improved scheme
combined with the modified pseudo-fermion action, compared with 
the leap-frog and the standard pseudo-fermion action.
For a fair comparison, we also have to take into account the overhead
caused by $S_{F1}$ of the modified pseudo-fermion action. Taking the 
total number of iterations needed by the solver within one trajectory we
arrive at a speed-up of $17000/6900=2.5$.
Based on this experience it would be very interesting to study the 
behaviour of an $O(\delta \tau^4)$ improved integration scheme.
Furthermore we find that
reversibility violations are of similar magnitude 
for the modified pseudo-fermion action and the standard pseudo-fermion action.
For the $8^3 \times 24$ lattice we performed extended runs with up to
8300 trajectories. From these runs we see that for the standard and the 
modified pseudo-fermion action as well as for different integration schemes,
the  autocorrelation times are the same, as long
as the acceptance rate is the same. 


The QCDSF collaboration \cite{QCDSF}  has simulated a $16^3 \times 32$ lattice
at $\beta=5.29$ with $c_{sw}=1.9192$ and $\kappa=0.1355$, which corresponds
to $m_{PS}/m_V\approx 0.7$. Preliminary results for 
a $24^3 \times 48$ lattice at $\beta=5.25$ with $c_{sw}=1.9603$ and 
$\kappa=0.13575$ are also available. This corresponds to 
$m_{PS}/m_V\approx 0.6$.
They have used the pseudo-fermion action
~(\ref{original}) and the leap-frog integration scheme with 
two time-scales~(\ref{ts}). They compared three versions of the 
algorithm: (A) The HMC with the standard pseudo-fermion action, (B)
the HMC with the modified pseudo-fermion action, where both $S_{F1}$ and 
$S_{F2}$ are on the same time-scale and (C) 
analogous to Sexton and Peardon \cite{Peardon}, 
$S_{F1}$  is on the same (smaller) time-scale as the gauge action. 
They have used
a chronologicial prediction of the start vector for the solver.

For $m_{PS}/m_V\approx 0.7$ they find  that (B) is about 2.8 times faster
than (A). Similar to our experience \cite{HaJa}, the step-size $\delta \tau$
of version (B) is twice the step size of (A). It is very surprising that an
other factor of 1.4 speed-up comes from the iteration number of the solver.
Naively one would expect that the larger step-size of version (B) would 
degrade the chronologicial prediction of the start vector and hence 
enlarge the iteration number. Maybe, with the modified pseudo-fermion action
the evolution of the solution vector becomes smoother and hence the 
chronologicial prediction more effective. With version (C), they even 
find a speed-up of about 3.4 compared with (A).
The preliminary results for $m_{PS}/m_V\approx 0.6$ are quite similar:
with version (B) the speed-up is about $3.5$ and with (C) $3.4$.  But here
the parameters of the algorithm are not yet tuned much. Again the 
speed-up is partially due to a reduction of the iteration number of the 
solver in the case of (B) and (C) compared with (A).

The ALPHA collaboration \cite{ALPHA} studied the pseudo-fermion 
action based on the splitting~(\ref{rainer}). They have used the 
partially improved 
integration scheme of Sexton and Weingarten. They  imposed
Schr\"odinger functional boundary conditions. They have studied 
lattices of size $L/a=6$ up to $L/a=24$ in a range $1.0 \le u \le 5.7$ of
their renormalized coupling, corresponding to  physical lattice sizes of  
$10^{-2}$ fm up to 1 fm. The range of $\beta$-values is $5.2 < \beta < 11.1$.

They have chosen the parameter $\rho$ of the splitting following the 
rule $\rho \approx (<\lambda_{min}>/<\lambda_{max}>)^{1/4}$, 
where $\lambda_{min}$ and $\lambda_{max}$
are the minimal and maximal eigenvalues of $M^{\dag} M$. The idea of this
choice is that both $\bar{M}$ and $W$ should have the same condition number.
The runs where performed on a APEmille computer using single precision 
floating point numbers. Therefore special attention on reversibility 
violations is needed. Their careful checks show that there is no relevant
difference in this respect between the two pseudo-fermion actions. To
check explicitly the effect of reversibility violations on the 
observables, they have varied the length of the trajectory. In addition to
the usual $\tau=1$ they studied $\tau=0.5, 1.5$ and $3$. 
They conclude that for $\tau=1$,  the reversibility violations do not 
invalidate the results of their simulations.

Since they performed rather long runs, even for $L/a=24$ they have 
generated 4000 trajectories, they can give meaningful results for the 
autocorrelation times. Measured in units of trajectories, for the same
acceptance rate, they find 
virtually no difference between the modified and standard pseudo-fermion 
action.

The main difference comes from the step-size that can be used to obtain 
an acceptance rate of about $80 \%$. 
For the modified pseudo-fermion 
action, the step-size
can be more than doubled compared with the standard case. There is  only 
a rather small increase of this factor towards smaller values of $\beta$ 
and larger $L/a$. 

As a further test of the modified pseudo-fermion action, 
I  have performed simulations at $m_{PS}/m_V \approx 0.4$ \cite{unpublished}. 
To this end I have taken the parameters reported by
Irving at the Lattice 2002 \cite{Irving}. He generated 
$2300$ trajectories for a
$16^3 \times 32$ lattice at  $\beta=5.2$, $c_{sw}=2.0171$,
$\kappa=0.1358$. He obtained $m_{PS}/m_{V}=0.43(2)$ and $m_{\pi} L=3.3(1)$,
where the last result indicates that the lattice size is actually too small 
to extract $L\rightarrow \infty$ results from the simulation. The result for 
the plaquette is 
$Plaq=0.53770(4)$ with an autocorrelation time of $\tau \approx 7$.
The step-size of  the leap-frog scheme is $\delta \tau = 1/400$.
The length of the trajectory is $\tau=1$.

In my run I have tested only the modified pseudo-fermion action based on 
eq.~(\ref{original}). The partially improved integration scheme of 
Sexton and Weingarten is used. 
First I equilibrated the system for the parameters chosen. After some rough 
search for the optimal value of $\rho$, I generated 
$50$ trajectories with 2 pseudo-fermion fields and the choice $\rho=0.05$.
With a step-size of $1/33$,  $P_{acc} = 0.81(4)$ is obtained. Taking into
account the fact that in the partially improved scheme the force has to be 
computed twice per step, we see a speed-up of $400/(2 \times 33) \approx 6$
compared with the run of Irving \cite{Irving}. 

Then I simulated with 3 pseudo-fermion fields, i.e. splitting the 
fermion matrix into three factors. The parameters
$\rho_1=0.4$ and $\rho_2=0.03$ of the corresponding pseudo-fermion action 
are guesses based on the experience with the two pseudo-fermion case.
The fermion matrix is normalized such that it is $1$ for $\kappa=0$.
After generating about 50 trajectories, we found that a larger step 
size than with  two pseudo-fermions could be reached. Based on that we
generated $1500$ trajectories with a step-size of $\delta \tau = 1/25$.
We find an acceptance rate of $P_{acc}= 0.809(6)$. The expectation 
value of the average plaquette is
$Plaq=0.53773(5)$, which is compatible with Irving's result \cite{Irving}.
The integrated auto-correlation time of the plaquette is $\tau=7.5\pm 3.0$.
This again confirms that auto-correlation times depend very
little on the pseudo-fermion action that is used.

In the simulation I have also seen a few spikes. The largest is 
$\Delta H \approx 3774$. The second largest is $\Delta H \approx 30$.
These spikes are rather small compared with those of ref. \cite{Namekawa}
which are of order $10^5$.  
Also $\langle \exp(-\Delta H) \rangle = 0.984(10)$ for my simulation 
seems to be fine. 
This result shows that the modified pseudo-fermion action not only
allows for a larger step-size but also tames the spikes.

\section{THE MULTIBOSON ALGORITHM}
The qq+q collaboration has recently reported results for two-flavour 
Wilson simulations at quark masses in the range 
$\frac{1}{6}m_s < m_{q} < 2 m_s$ using the TSMB version of the algorithm. 
A discussion of the  performance, depending on the quark mass, the lattice size
and the lattice spacing, is given 
in \cite{Montvay} (and refs. therein).
The dependence on the pseudo scalar mass is given by
$cost \propto m_{PS}^{-z}$ with $3 < z < 4$, which seems a little worse than  
the HMC algorithm.
This is actually not too surprising. The increase of CPU cost with decreasing
sea-quark mass in the 
MB algorithm and variants of it have mainly two sources:  First the increase
of the order of the polynomial and second the increase of autocorrelation 
times that are related to the order of the 
polynomial. The increase of the order of the polynomial should be similar
to the increase of the number of iteration of the solver 
in the HMC algorithm, which is the major source for the increasing effort
in the HMC algorithm. 
On the other hand, for the MB algorithm, 
the numerical costs increase trivially with the 
lattice volume: $cost \propto (L/a)^4$, which compares favourably with the 
HMC algorithm, 
where we have $cost \propto (L/a)^5$ for the leap-frog integration 
scheme. 
The qualitative 
disadvantages of the MB algorithm compared with the HMC algorithm are 
the large memory requirement to store the boson-fields and the difficulty 
to implement the algorithm for improved fermion actions.

\section{CONCLUSIONS}
Tuned once more,
the HMC algorithm seems to be a good choice for the simulation
of dynamical Wilson fermions.
One should 
take great care of all details like the precise choice of the 
integration scheme and the preconditioning of  
the fermion matrix.  
The modified pseudo-fermion action \cite{MHschwinger}
speeds up the HMC simulation by allowing for a larger step-size of the 
integration scheme. Higher order 
integration schemes seem to profit more from the modified pseudo-fermion action
than the leap-frog scheme. 
Hence, it seems likely that the scaling of the cost with the volume of 
the lattice
can be improved to $cost \propto (L/a)^{4.5}$ by using an $O(\delta \tau^4)$ 
integration scheme, without negative side effects on the scaling with 
$m_{PS}$.
The modified pseudo-fermion action 
introduces no unwanted side-effects like larger violations of reversibility
or larger autocorrelation times.
The proposal of Peardon \cite{Peardon} is more flexible than that of ref.
\cite{MHschwinger}. Hence, it might lead to even larger speed-ups. 
Also it is
straight forward to apply \cite{Peardon} to the Polynomial HMC algorithm.
Whether new ideas \cite{Luescher} will beat the HMC remains to be seen.

%

\end{document}